\documentclass[twocolumn]{article}

\usepackage[english]{babel}
\usepackage[utf8x]{inputenc}
\usepackage[T1]{fontenc}
\usepackage{amsfonts}
\usepackage{mathrsfs}
\usepackage{mathtools}
\usepackage{subcaption}
\usepackage{caption}
\usepackage{graphicx}
\usepackage[a4paper,top=2cm,bottom=2cm,left=1.5cm,right=1.5cm]{geometry}

\usepackage{amsmath}
\usepackage{graphicx}
\usepackage[colorinlistoftodos]{todonotes}
\usepackage[colorlinks=true, allcolors=blue]{hyperref}
\usepackage{float}
\usepackage[ruled,linesnumbered]{algorithm2e}
\usepackage{algpseudocode}
\title{
	\usefont{OT1}{bch}{b}{n}
	\normalfont \normalsize \textsc{} \\ [8pt]
	\huge 
	A Conditional-Probability-Distribution Model for Bandwidth Estimation with 
	Application in Live Video Streaming\\
}
\usepackage{authblk}
\author[1]{Weijia Zheng}
\affil[1]{Department of Information Engineering, CUHK}
\affil[1]{Department of Mathematics, CUHK}
\affil[1]{
wjzheng@link.cuhk.edu.hk}

\date{}
\begin{document}

\maketitle

\begin{abstract}
Experience of live video streaming can be improved 
if the video uploader have more accurate knowledge 
about the future available bandwidth. 
Because with such knowledge, one is able to know 
what sizes should he encode the frames to be in 
an ever-changing network. 
Researchers have developed some 
algorithms to predict throughputs in the literature, 
from where some simple hence practical ones.
However, limitation remains as most current bandwidth prediction 
methods are predicting a value, or a point estimate, of future bandwidth. 
Because in many practical scenarios, 
it is desirable to control the performance to some targets, 
e.g., 
video delivery rate over a given target percentage, 
which cannot be easily achieved via most current methods.

In this work, we propose the use of probability distribution to model future bandwidth. 
Specifically, we model future bandwidth using past data transfer measurements, 
and then derive a probability model for use in the application.
This changes the selection of parameters in application into a probabilistic manner 
such that given target performance can be achieved in the long run.
Inside our model,
we use the conditional-probability method
to correlate past and future bandwidth and hence further improve the 
estimating performance.

\end{abstract}

~\

{\textbf{Keywords} \\
Network throughput estimation, conditional probability, 
relative frequency, live video streaming, empirical method}

\section{Introduction}

With the development of smartphones and 
high-speed mobile data networks such as 3G and 4G/LTE, 
live video streaming has long become part of our lives 
in the entertainment or casual fields. 
In recent years, it has played a vital role in the workplace as well. 
As we all experienced in person, 
the year of 2020 witnessed Zoom's significant increase [1] 
in the usage of remote work, 
distance education, and online social relations. 
 
Given the importance of video streaming and 
the high peak bandwidth nowadays in mobile data networks, 
bandwidth fluctuation remains a challenging obligation 
for its unpredictable characteristic by its wireless nature, 
which may affect the quality of clients' experience. 
Several prediction methods were established or used in the literature, 
such as arithmetic mean (AM), multiple linear regression (MLR), 
ARIMA, LSTM, etc. However, most of them predict a point estimate 
of future bandwidth, or gives a confidence interval with further assumptions,
which is hard to guarantee its accuracy to a specified level in practice.

Motivated by the ideas of using an empirical conditional probability 
for prediction in financial [2] and transportation fields [3], 
this work contributes to establishing a conditional-probability model in bandwidth prediction.
And a simulator of the uploading part of a live video streaming, 
referencing the one in [4], 
is implemented for simulation
and demonstrating the feasibility of such a method.  
To mimic a real network environment, the simulator uses a 
packet-level with timestamp trace data measured 
from real-world network sources of 3HK 4G. 

The rest of the paper is organized as follows. 
Section 2 introduces the settings of the problem. 
Some analyses and related works are 
presented in Section 3.
In Section 4 we will derive the proposed encoding scheme in detail. 
Numerical results of it on different network environments 
are shown in Section 5, 
before conclusions are outlined in Section 6.

Besides, unless otherwise stated, all the bitrate variables are in Mbps, 
and all the time variables are in seconds in this paper.

\section{Problem Background}
In our study, to concentrate on the uplink part, 
we considered the uplink part exclusively with some further simplifications.
We do not consider the downlink part of the streaming process in the simulator 
and the general scenario can be described as the following.

An uploader generates frames one by one 
with an equal time difference (i.e., 1/FPS second) 
in between. Assume that TCP is used, 
once a frame is sent from the uploader side, 
the uploader will not consider it anymore but 
will only take care of later frames.
The uploader then starts transmitting 
some newest possible frame as buffer time allows. 
When the streaming process terminates, 
should there be any not-ever-sent frame, 
we count it as a frame loss. 

\begin{figure}[H]
	\centering
	 \includegraphics[width=0.48\textwidth]{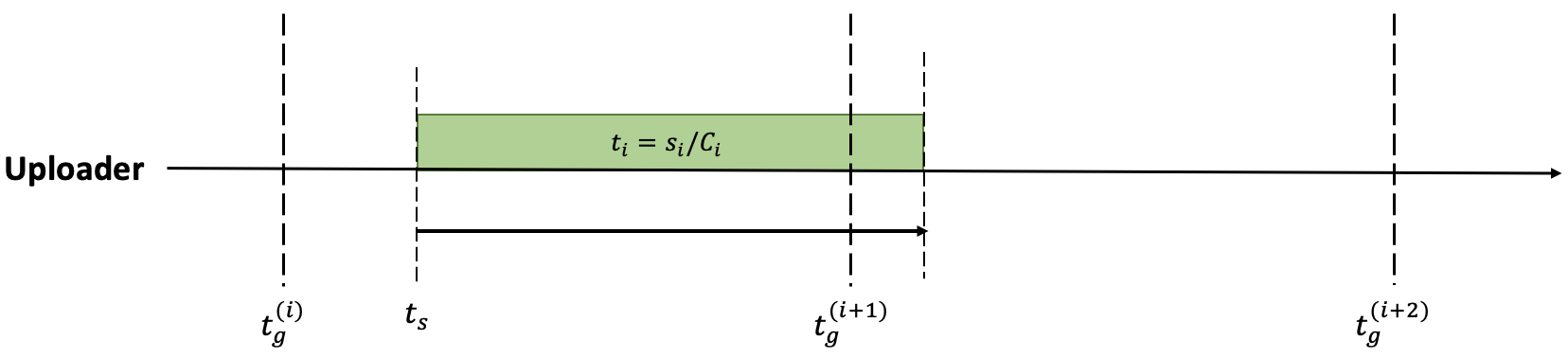}
	 \caption{Frame sending timeline}
\end{figure}

Denote a frame $f_i$'s generating time as $t_g^{(i)}$.
Once the uploader finishes transmitting a frame $f_i$ 
whose time cost is $t_i$, 
it can calculate the (just now) average throughput in the past 
$t_i$ time interval $C_i$ using 
$$C_i = s_i / t_i,$$ 
where $s_i$ is frame $f_i$’s size 
which is determined by the uploader itself.
And $t_i$ is how much time the uploader cost to process $f_i$, 
which the uploader can always make a record on. 
When frame $f_{i+1}$ needs to be transmitted, 
the uploader may predict how will the network soon
(i.e., $C_{i+1}$) behave 
through utilizing all those pieces of knowledge about the network's history 
it was obtained using the above way.

A study carried by Zou et al. [5] revealed that 
the optimal case happens when we can achieve 
$$C_{i+1} = \widehat{C_{i+1}}.$$ 
Because then $\forall j,$ we can let $f_j$ 
finish its transmission at $t_g^{(j+1)}$, the exact time 
when frame $f_{j+1}$ is just generated. 
By this, we are sending $s_{j}$ as large as possible 
while not losing any single frame and introducing no delay. 

Besides, a minimal frame size $s_{\text{min}}$, namely,
$\forall j, s_j\geq s_{\text{min}}$
is required.
Because we cannot let $s_j$ be arbitrarily small. 
We also set some fixed value $t_B$ as our initial buffer time, 
which usually is some multiple of $1/\text{FPS}$. 
This allows the uploader to transmit some older frames 
that are still within the buffer time. 
Note that the buffer time may change with $i$, 
since it may be eaten up by overestimating the throughputs in the past, 
which causes a longer transmission time than expected.

\section{Related Works}
We use two simple metrics to measure 
how an encoding scheme performs during a time period of length $\mathcal{T}$ (in second). 
They are 
$$\text{loss rate} \coloneqq 
\frac{\text{$\sharp$ frames discarded during the $\mathcal{T}$ period}}
{ \text{$\sharp$ frames generated during the $\mathcal{T}$ period} }$$ 
and 
$$\text{(average) bitrate} \coloneqq  \frac{1}{\mathcal{T}}\sum_{i} s_i.$$ 
An ideal algorithm will lead to low (close to zero) loss rate 
and large average bitrate. 
Both can be reached at the same time 
when the perfect prediction mentioned is achieved. 

It is noticeable that the two metrics (loss rate and bitrate) 
can be easily optimized individually.
By letting $s_{\text{min}}$ to be very large or small values 
and encode every frame as size $s_{\text{min}}.$
This gives us an idea that to eliminate loss rate we 
do not necessarily have accurate knowledge of $C_j$'s, but some lower bounds will do.
And there exists an upper limit of bitrate which we can use to check how an algorithm behaves.

Besides sophisticated methods explored in [6], [7] and [8],
two seemly-trivial throughput predicting algorithms got suggested in [9]. 
They are arithmetic mean (AM)
and multiple linear regression (MLR). 
The AM algorithm returns a moving average of the past $C_i$'s. 
Suppose $K$ is the number of past $C_i$'s we look back on, 
then one calculates
\begin{equation}
	\widehat{C_{i+1}} = \frac{1}{K}\sum_{k=1}^{K}C_{i+1-k}
\end{equation}
as an estimate for $C_{i+1}$. 
However, this assumes each $t_{i+1-k}$'s are the same, 
which cannot be guaranteed in reality, 
hence the (1) should be adapted to 
\begin{equation}
	\widehat{C_{i+1}}=
\frac{\sum_{k=1}^{K} C_{i+1-k}\cdot t_{i+1-k} }{\sum_{k=1}^{K}t_{i+1-k}}.
\end{equation}

Multiple linear regression has a general form 
(one may manipulate the formula to avoid multicolinearity)
$$\widehat{C_{i+1}}=b_0+b_1C_i+b_2C_{i-1}+...b_KC_{i-k+1},$$
where $b_0$ is the intercept and $b_k$'s are weights. 
The famous ARIMA, discussed in [7] and [8], and many others are variants of MLR.
However, note that this also needs to be adapted in practice, 
since $t_{i-k+1}$'s may vary.
These algorithms do not require much training and are relatively simple to apply,
hence got suggested by many.

As one may be expected, in [6] and [8] LSTM also got some praise.
However, as different people are using different datasets, and the results between
different algorithms are quite matched in strength,
it is hard to say there is a best algorithm or a game-changer in this topic.

\section{Our Proposed Method}
\subsection{Motivation}
For any fixed given time $t$, we look back on the past and divide the past time into 
$\tau$ equally sized non-overlapping intervals, 
i.e., $(t-\tau,t],(t-2\tau,t-\tau],...,\big( t-(j+1)\tau,t-j\tau \big]...$
where $j\in \mathbb{N}.$

\begin{figure}[H]
	%   \centering
	  \includegraphics[width=0.48\textwidth]{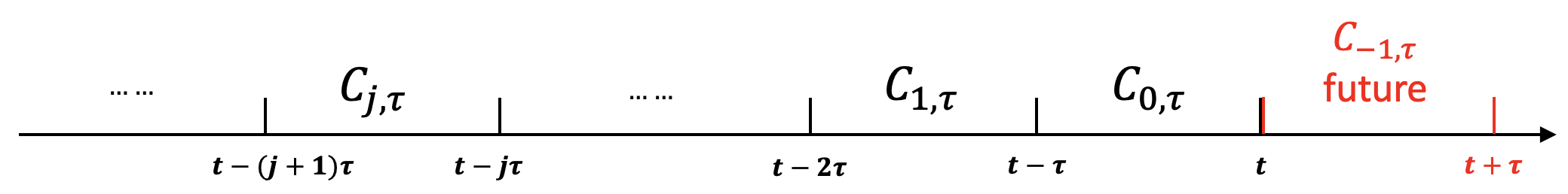}
	  \caption{Depiction of the timeline}
\end{figure}

Denote $C_{j,\tau}$ as the average throughput in the time interval 
$\big( t-(j+1)\tau,t-j\tau \big)$. 
In other words, 
the previous $j$-th time interval of length $\tau$ counting from $t.$
Since the throughput series indexed by $j$, 
i.e., $\{C_{j,\tau} | j\in \mathbb{N} \}$ 
forms a stochastic process over time.
It is conceivably to believe that the throughput is highly correlated with its past values
during some past time intervals. Especially, 
we may expect more recent past throughput data to exhibit higher correlation 
with future throughput and is therefore more helpful in predicting the (short-term) future. 
Which is verified by [9] that for some fixed $\tau=5$, 
$I(C_{0,\tau} ; C_{k,\tau} ) $ decreases as $k\in \mathbb{N}$ increases. 

Motivated by the research works in predicting short term traffic speed, 
which utilize history data to generate a Markovian state transition matrix [10] or a 
conditional CDF [3] to predict future travel speed, we propose a conditional PDF based 
method that is able to predict short-term future throughput as well as control 
the streaming behaviour in terms of frame loss rate at the uploader side.

\subsection{Implementation Procedures}
To stay the loss rate low meanwhile not wasting 
too much available bandwidths, 
we try to find promising $s_i$'s.

Suppose we want the loss rate to be no larger than some fixed $\epsilon \in (0,1).$
Now is time $t_s^{(i)}$, frame $f_i$ starts to transmit.
By discussions in section 2, $f_i$ will be discarded if 
\begin{equation}
	t_s^{(i)}+s_i/C_i>t_{g}^{(i+1)}+t_{b}^{(i)}.
\end{equation}
Recall that $C_i$ is defined to be the average throughput during $f_i$'s transmission.
$t_{b}^{(i)}(\neq t_B \text{ in general})$ 
is the available buffer time at this moment.

Let the probability of $f_i$ not to be discarded be no less than $1-\epsilon$.
Then we obtain
$$P(t_s^{(i)}+s_i/C_i \leq t_{g}^{(i+1)}+t_{b}^{(i)}) \geq 1-\epsilon,$$ 
which is equivalent to 
\begin{equation}
	P \big(C_i(t_{b}^{(i)}+T_i) \geq s_i \big) \geq 1-\epsilon,
\end{equation}
where $$T_i \coloneqq [t_{g}^{(i+1)}-t_s^{(i)}]^+.$$ 
In order to enhance the bandwidth efficiency, we choose an $s_i$ 
such that the equality in (4) is taken, i.e., 
\begin{equation}
	P \big(C_i(t_{b}^{(i)}+T_i) \geq s_i \big) = 1-\epsilon.
\end{equation}

Since the uploader keeps a log of past throughputs of the network, 
it is possible to construct a time series 
\begin{equation}
	\{C_{j,t_{b}^{(i)}+T_i}\cdot (t_{b}^{(i)}+T_i) \mid j=0,1,...\}
\end{equation}
Where $ C_{j,t_{b}^{(i)}+T_i}$ 
denotes the average throughput in interval 
$\Big(t_s^{(i)}-(j+1)(t_{b}^{(i)}+T_i), t_s^{(i)}-j(t_{b}^{(i)}+T_i)\Big].$
That is, the uploader constructs an array $\hat{\textbf{s}}$, whose $j$-th entry 
is how many Mb of data can be transmitted in the $j$-th interval with length
$(t_{b}^{(i)}+T_i)$ counting back from $t_s^{(i)}$ following ${\textbf{t}}$.

Define $s_{k,\tau} \coloneqq C_{k,\tau}\cdot \tau$.
(5) then changes to 
\begin{equation}
	P \big( s_{-1,t_{b}^{(i)}+T_i} \geq s_i \big) = 1-\epsilon.
\end{equation}
According to (6) and the discussions following it,
the uploader maintains an array 
$$\hat{\textbf{s}} \coloneqq[s_{0,t_{b}^{(i)}+T_i},
s_{1,t_{b}^{(i)}+T_i},...,s_{J-1,t_{b}^{(i)}+T_i}]$$
as a time-series, which represents 
a relative frequency of $s_{-1,t_{b}^{(i)}+T_i}$ in terms of its realizations in (7). 
However, to reach (7), we let the probability be conditioning on its predecessor,
$s_{0,t_{b}^{(i)}+T_i}$, 
which is known as that's the first element of $\hat{\textbf{s}}.$ 
Thus (7) can be replaced by 
\begin{equation}
	P \big( s_{-1,t_{b}^{(i)}+T_i} \geq s_i \mid \underbrace{ s_{0,t_{b}^{(i)}+T_i}}_{\text{some known value}} \big) = 1-\epsilon.
\end{equation}

Define 
$$\hat{\textbf{s}_{|0}}\coloneqq [s_{j-1,t_{b}^{(i)}+T_i}\in\hat{\textbf{s}} 
\text{ for } s_{j,t_{b}^{(i)}+T_i} \approx s_{0,t_{b}^{(i)}+T_i} ]$$ 
Remember that $s_{-1,t_{b}^{(i)}+T_i}=C_{-1,t_{b}^{(i)}+T_i}\cdot (t_{b}^{(i)}+T_i)$ 
is a proper size for $f_i$, which 
we want to determine by estimating $C_{-1,t_{b}^{(i)}+T_i} 
\Big( \text{plays the role as $C_i$ } \text{in (4)} \Big)$. 
Hence $P(s_{-1,t_{b}^{(i)}+T_i}|s_{0,t_{b}^{(i)}+T_i})$ 
can be approximated by $\hat{\textbf{s}_{|0}}$ since 
the $\hat{\textbf{s}_{|0}}$ is its one conditional relative frequency.

By (8), one should take 
\begin{equation}
	s_i = Q_{s_{-1,t_{b}^{(i)}+T_i}|s_{0,t_{b}^{(i)}+T_i}}(\epsilon) 
\approx \textbf{quantile}(\hat{\textbf{s}_{|0}},\epsilon)
\end{equation}
as our chosen frame size $s_i$ for $f_i,$
where $Q_X(\gamma)$ denotes the $\gamma$ quantile of random variable
$X$. And $\textbf{quantile}(\vec{a},\gamma)$ is the numpy quantile function in Python, 
which returns the value at the $\gamma$-th quantile value of $\vec{a}$. 

The algorithm of deciding $s_i$ is articulated in Algorithm 1.

\begin{algorithm}[t]
	\caption{Determination scheme of $s_i$}
	\KwData{
	
	updated frame-level sizes: $\textbf{s}=[s_{0},s_{1},...,s_{k}]$,\\
	updated frame-level timestamp: $\textbf{t}=[t_{0},t_{1},...,t_{k}]$,\\
	buffer time at the moment, $t_{b}^{(i)}$, \\
	time before $t_{g}^{(i+1)}$, $T_i=[t_g^{(i+1)}-t_s^{(i)}]^+$,\\
	how many intervals do we trace back, $J \in \mathbb{N^*}$, \\
	loss rate target, $\epsilon \in (0,1)$.
	}
	\KwResult{Maximal frame size $s_i$ s.t. (5) holds.}
	\BlankLine
	\begin{algorithmic}
		\State $\hat{\textbf{s}} \gets [~]$\
		\State $\text{j} \gets 0$\

		\While{$\text{Len}(\hat{\textbf{s}})<J$}
		{
			$\hat{\textbf{s}}$ $\textbf{append}
			\bigg(\textbf{F} (j)-
			\textbf{F} (j)  \bigg)$\
			$j \gets j + 1 $\
		}
		
		% \While(){\Call{Len}{$\vec{v}$}$<J$ } {
		% 	$\vec{v}$ $\textbf{append}\bigg(\text{F} (t_F)-\text{F} \Big(t_F-(t_{b,i}+ T_i) \Big) \bigg)$ \\
		% \CommentSty{(take $\alpha=\beta=1$ for simplicity)}
		% \EndWhile}

		\State $\hat{\textbf{s}}_{\mid 0} \gets$ 
		$\big[\hat{\textbf{s}}[n-1] \in \hat{\textbf{s}} 
		\textbf{  for  } \hat{\textbf{s}}[0] \approx \hat{\textbf{s}}[n] \big]$ \\ 
		\CommentSty{($\approx$ can be defined as within $0.05$ variation)}
		\State $s_i \gets \textbf{quantile}( \hat{\textbf{s}}_{\mid 0} ,\epsilon)$ \\
		\Return{$s_i$} 

		\hfil

		% t_{b}^{(i)} +T_i
		\Procedure{\textbf{F}}{$j$}
			\State $U \gets \max\{u \mid \sum_{m=u}^{k}t_m \geq j\cdot(t_{b}^{(i)} +T_i) \}$
			\State $t_\text{Res,U} \gets \sum_{m=U}^{k}t_m - j(t_{b}^{(i)} +T_i)$
			\State $L \gets \max\{l \mid \sum_{m=l}^{U-1}t_m \geq t_{b}^{(i)} +T_i- t_\text{Res,U} \}$
			\State $t_{\text{Res,L}} \gets t_{b}^{(i)} +T_i -t_{\text{Res,U}}-\sum_{m=L+1}^{U-1}t_m$

			\State $F_j \gets 
			\frac{t_\text{Res,U}}{t_U}\cdot s_U + 
			\frac{t_\text{Res,L}}{t_L}\cdot s_L + 
			\sum_{m=L+1}^{U-1} s_m$ 

			\State \textbf{return} $F_j$ 
		\EndProcedure
	
	\end{algorithmic}
\end{algorithm}

After $s_i$ is determined, then by the end of $f_i$'s transmission, 
$t_i$ will be available.
Then we update the frame-level history data $\textbf{s}$ and $\textbf{t}$ by appending
$s_i$ and $t_i$ respectively, and update the 
buffer time based on if $t_i\leq 1/\text{FPS}$ or $t_i > 1/\text{FPS}$. 
Thus we are ready for the next round of transmission.

One may note that in (4), to be more conservative, one can change $t_b^{(i)}$ by any smaller
value $\alpha t_{b}^{(i)}$ with $\alpha \leq 1.$  
And then all the following steps should change correspondingly.

\section{Simulation Results}
The following section shows the results of the proposed method. 
The explanatory variables are the two metrics, 
minimal frame size $s_{\text{min}}$ and 
initial buffer time $t_B.$ 
We ran simulations on two different networks. 
For simplicity, we will only compare with AM algorithm, 
with its parameter in (2), $K=5,16,128$ respectively. 
And we always let the target loss rate to be $\epsilon=0.05.$
Forall plots whose x-axis is $t_B$, 
we let the $s_{\text{min}}$ fixed to be some small value.
Since in a live video streaming, 
the $t_B$ cannot set be too large, we let it be up to $0.1$s.
Similarly, too large loss rate like $20\%$ or above are 
not meaningful, we will focus on when the loss rate is below $20\%.$

Finally, the training time for our algorithm is $120$ seconds, 
and the $\text{FPS}=60.$

The below groups of figures show how our conditional probability method 
behaves in terms of loss rate and bitrate when $s_{\text{min}}$ and $t_B$ increases individually. 
The network (call it network 1) in Fig. 3 and Fig. 4 has a mean of throughput 
$\approx 12 \text{ Mbps}$. 
And the network (network 2) in Fig. 5 and 6 has a mean of throughput  $ \approx 8 \text{ Mbps}$.
We compare with AM algorithms and the marginal version (does not use conditioning) of our method. 

\begin{figure}[H]
		% \centering
		\subfloat{
			\label{ref_label1}
			\hspace*{-0.4in}
			\includegraphics[width=5cm,height=5cm]{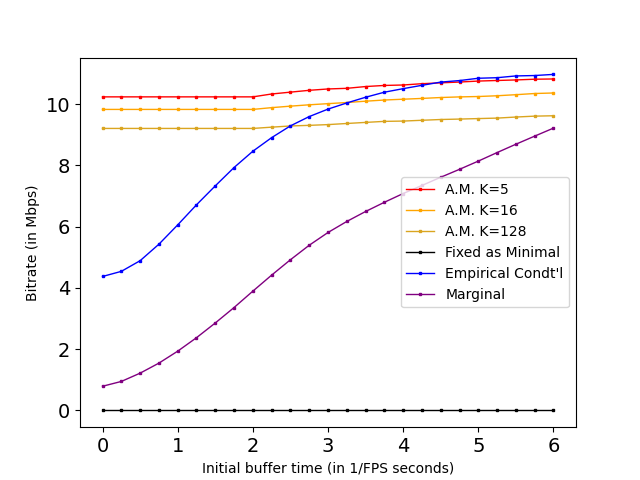}
			}
		\subfloat{
			\label{ref_label2}
			\hspace*{-0.3in}
			\includegraphics[width=5cm,height=5cm]{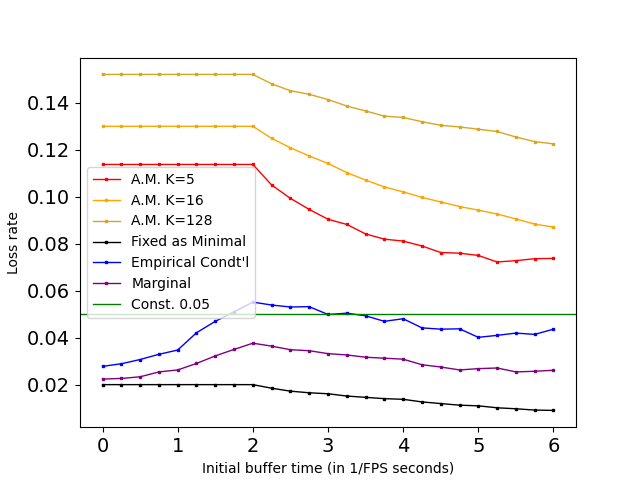}
			}\\
		\vspace{-1\baselineskip}
		\caption{Network 1. Fix $s_{\text{min}}$ be small, bitrates and loss rates as $t_B$ increases.}
		\label{ref_label_overall}
\end{figure}

\begin{figure}[H]
	% \centering
	\subfloat{
		\label{ref_label1}
		\hspace*{-0.2in}
		\includegraphics[width=5cm,height=5cm]{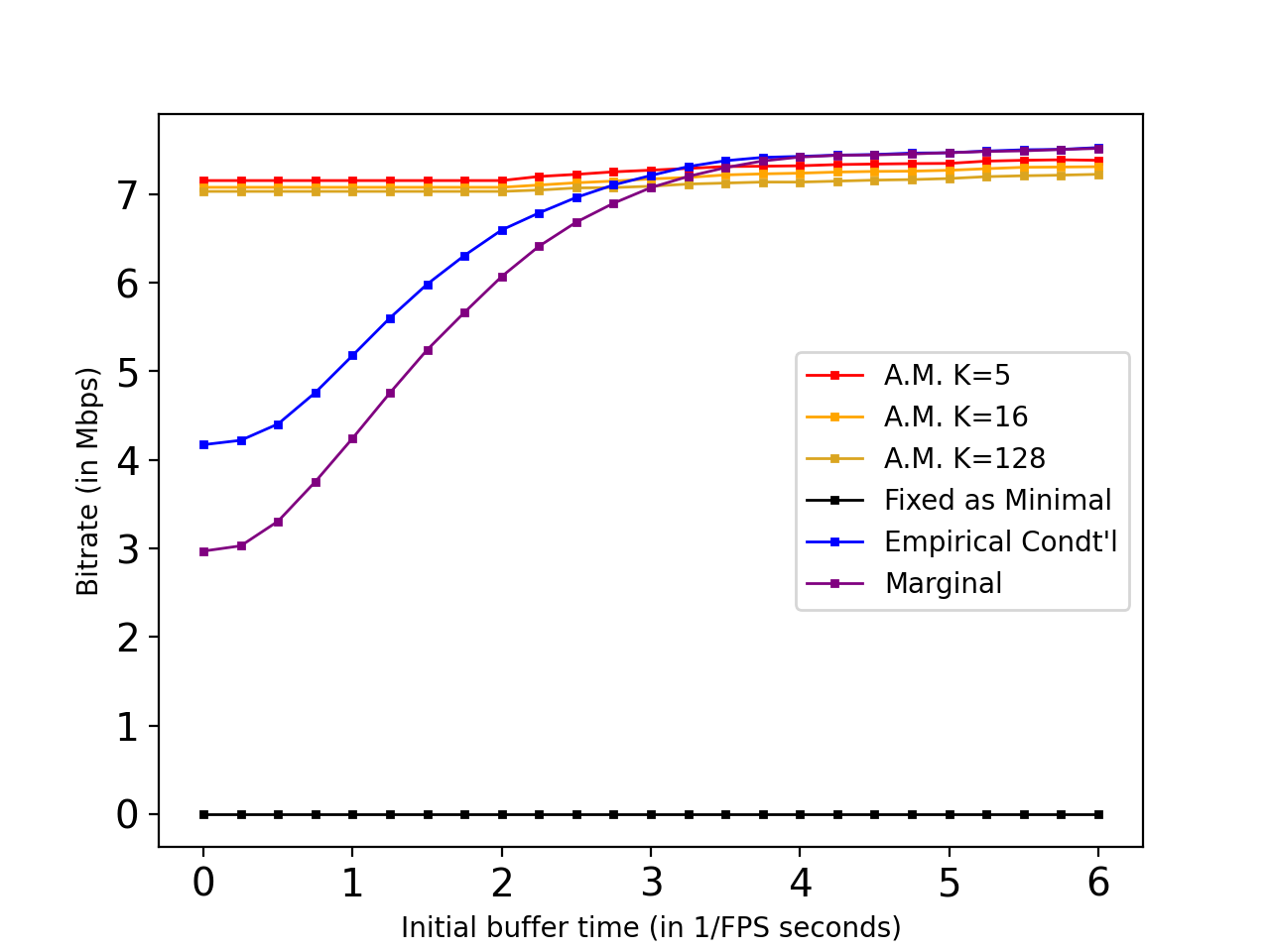}
		}
	\subfloat{
		\label{ref_label2}
		\hspace*{-0.2in}
		\includegraphics[width=5cm,height=5cm]{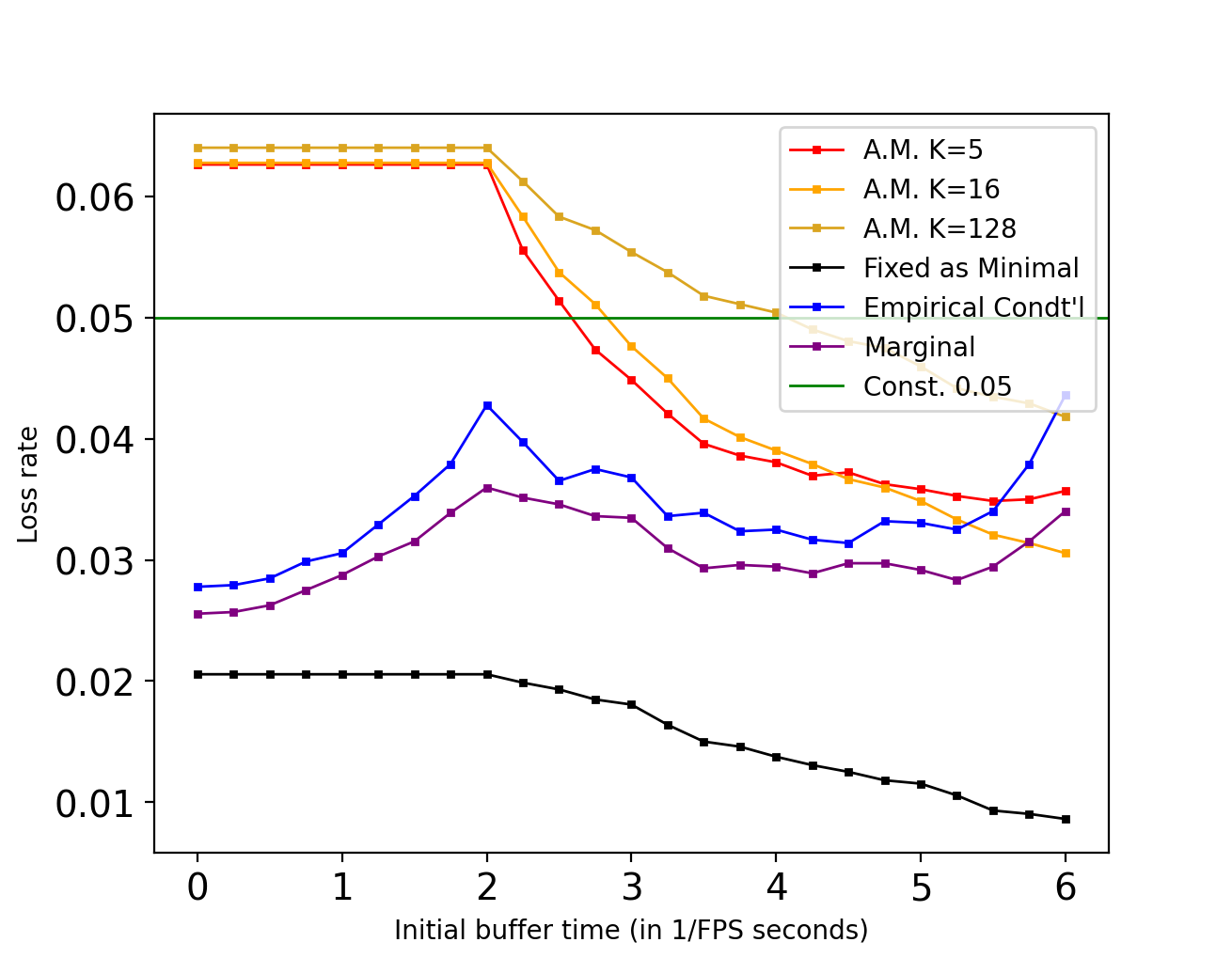}
		}\\
	\vspace{-1\baselineskip}
	\caption{Network 2. Fix $s_{\text{min}}$ be small, bitrates and loss rates as $t_B$ increases.}
	\label{ref_label_overall}
\end{figure}

Note that when $t_B$ is small, our method gives lower bitrate comparing
with the AM algorithms, which is reasonable since that is the trade off 
for having a controlably (low) loss rate. Because AM algorithms are 
mean-bitrate pursuing algorithms and ours is not.
As $t_B$ increases, our method's 
bitrate increases while the loss rate remains fluctuating around $0.05.$
When $t_B$ reaches $0.1$s ($=6/\text{FPS}$), its bitrate is leveraged to the maximal level. 
Which is just as what we expectated.

\begin{figure}[H]
	% \centering
	\subfloat{
		\label{ref_label1}
		\hspace*{-0.2in}
		\includegraphics[width=5cm,height=5cm]{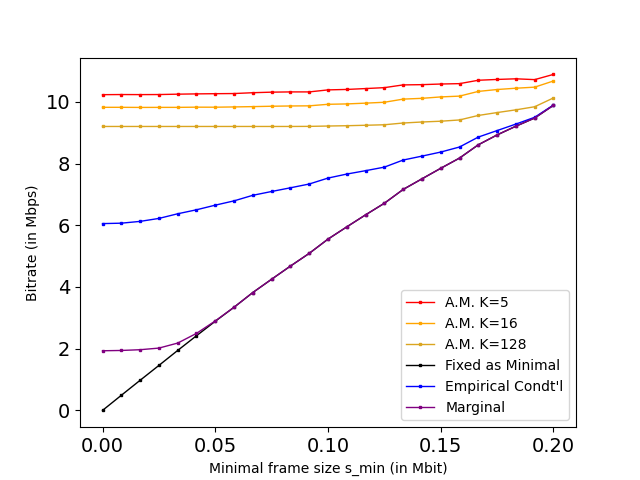}
		}
	\subfloat{
		\label{ref_label2}
		\hspace*{-0.2in}
		\includegraphics[width=5cm,height=5cm]{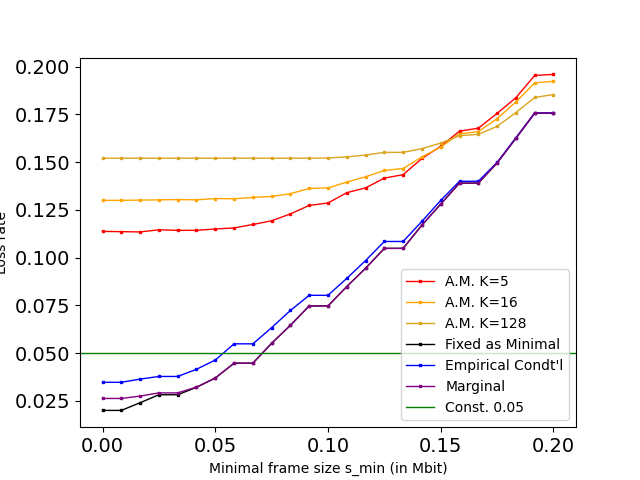}
		}\\
	\vspace{-1\baselineskip}
	\caption{Fix $t_B=1/\text{FPS}$, bitrates and loss rates as $s_{\text{min}}$ increases.}
	\label{ref_label_overall}
\end{figure}

\begin{figure}[H]
	% \centering
	\subfloat{
		\label{ref_label1}
		\hspace*{-0.3in}
		\includegraphics[width=5cm,height=5cm]{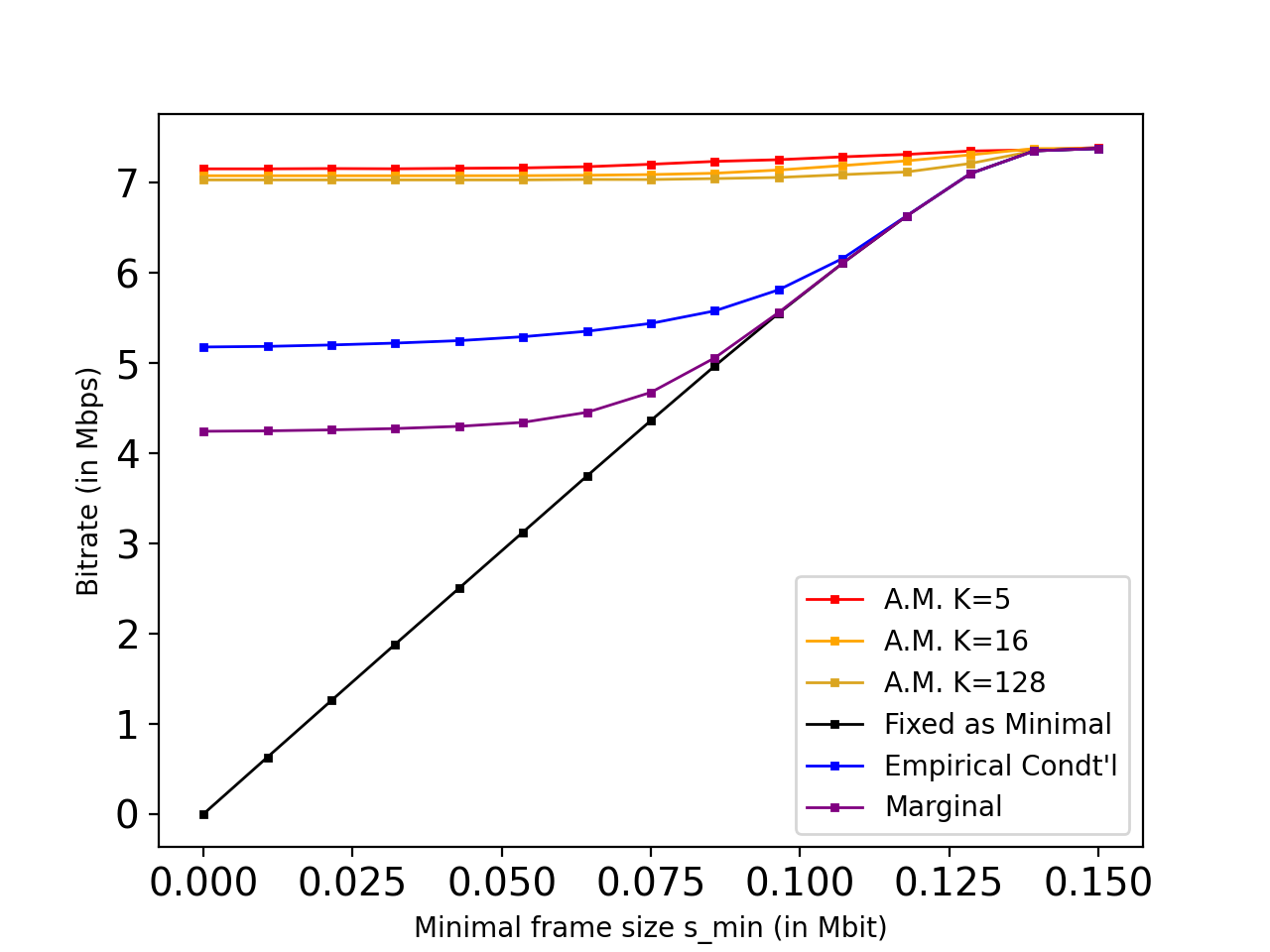}
		}
	\subfloat{
		\label{ref_label2}
		\hspace*{-0.3in}
		\includegraphics[width=5cm,height=5cm]{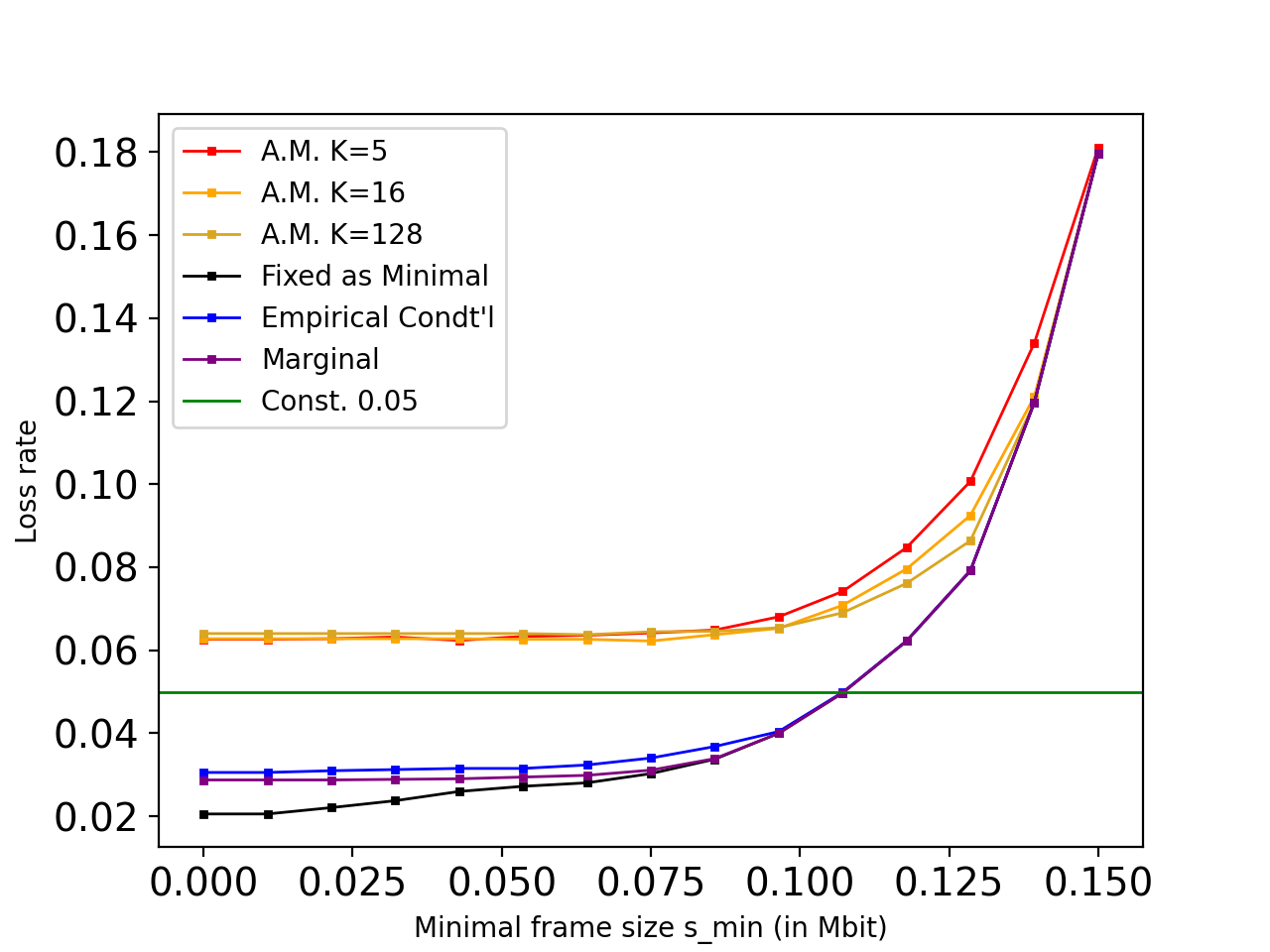}
		}\\
	\vspace{-1\baselineskip}
	\caption{Fix $s_{\text{min}}$ be small, bitrates and loss rates as $t_B$ increases.}
	\label{ref_label_overall}
\end{figure}

The above Fig. 5 and 6 shows when fix $t_B$ to be $1/FPS$, 
how bitrate and loss rate changes as $s_{\text{min}}$ 
increases in network 1 and 2 respectively. 
As expected again, our method suffers a bitrate loss comparing with 
the AMs as the trade off for controling the loss rate to be under 
$\epsilon=0.05,$ which can be observed as a flat line in the plots.
However, it should be surprising comparing with the minimal frame scheme
and the marginal probability method's behaviour. 
Conditional probability method is really reaching some balance 
between loss rate and bitrate, under any given target $\epsilon$ and 
buffer time $t_B.$
When the $s_{\text{min}}$ goes large, then all algorithms 
will be taken over by $s_{\text{min}}$. That is why all algorithms 
"converges" together in the right end of Fig. 5 and 6. 

From the above results, the conditional probability method 
is able to control certain loss rate target, 
while its bandwidth efficiency can be relatively high.

\section{Conclusion}
This work proposed and demonstrated a history-based short-term bandwidth
predicting method called conditional probability method, 
which especially has a practical value in 
live video streaming.

Rather than doing a (point) estimation on the future throughput, 
conditional probability method uses the 
empirical conditional relative frequency to find the largest frame size 
while eliminating the probability of frame loss under any networks. 
It is expected to give good performances in reality via 
its simulation results under different packet-level testing sets.

Possible improvements such as giving a more concrete theoretical proof 
of this method, a faster implementation of 
the algorithm in both space and time complexities, 
and ways to obtain more reliable past data are expected in future works 
to further enhance this method's efficiency and robustness.

\end{document}